# Resolution Improvement for Optical Coherence Tomography based on Sparse Continuous Deconvolution

**ZHENGYU QIAO[1], YONG HUANG[1,2] AND QUN HAO[1,3]**

[1] *School of Optics and Photonics, Beijing Institute of Technology, No.5 South Zhongguancun Street, Haidian, Beijing 100081, China*
[2] *huangyong2015@bit.edu.cn*
[3] *qhao@bit.edu.cn*

**Abstract:** We propose an image resolution improvement method for optical coherence tomography (OCT) based on sparse continuous deconvolution. Traditional deconvolution techniques such as Lucy-Richardson deconvolution suffers from the artifact convergence problem after a small number of iterations, which brings limitation to practical applications. In this work, we take advantage of the prior knowledge about the sample sparsity and continuity to constrain the deconvolution iteration. Sparsity is used to achieve the resolution improvement through the resolution preserving regularization term. And the continuity based on the correlation of the grayscale values in different directions is introduced to mitigate excessive image sparsity and noise reduction through the continuity regularization term. The Bregman splitting technique is then used to solve the resulting optimization problem. Both the numerical simulation study and experimental study on phantoms and biological samples show that our method can suppress artefacts of traditional deconvolution techniques effectively. Meanwhile, clear resolution improvement is demonstrated. It achieved nearly twofold resolution improvement for phantom beads image that can be quantitatively evaluated.

## 1.  Introduction

Optical coherence tomography (OCT) is an optical imaging technique that can provide non-invasive, cross-sectional imaging of biological tissue with micrometer spatial resolution [1-2]. While micrometer resolution is already relatively high for biological tissue imaging, the demand of higher resolution is consistently present [3-5]. Further resolution improvement in OCT images can reveal unseen microstructures that will contribute to accurate diagnosis [6,7].

Theoretically, axial resolution of OCT image is limited by the full-width-half-maximum (FWHM) of the light source's coherence function, which is inversely proportional to its spectral bandwidth [8,9]. To date, multiple groups have demonstrated OCT with about 1 μm axial resolution [3], where light source with an extra-large bandwidth of about 300 nm were used [10,11]. However, light sources with an extra-large bandwidth are difficult to build, and their implementation also made the system optical design complicate. Moreover, pushing the axial resolution beyond this limit through further expansion on the bandwidth could be very challenging: it will require not only the technological progress in laser sources but also proper handling of the collateral chromatic aberration and dispersion [12,13].

Lateral resolution is determined by the numerical aperture of the objective lens [8,9], which usually suffers from the tradeoff with depth-of-focus [14-16]. Previous methods addressing this tradeoff between the lateral resolution and depth-of-focus included multi focusing [16], optical light beam shaping [17,18] and dynamic focusing [19]. Although all these methods accomplished the lateral resolution improvement, it still required modification of the hardware.

Therefore, numerical methods were proposed to boost resolution of OCT images without hardware modifications. The most noteworthy examples are spectral estimation methods, deconvolution methods and deep learning methods.

Autoregressive spectral estimation technique alternative to the discrete Fourier transform has been used to improve the axial resolution [20-22]. Although spectral estimation OCT has shown promising results, there are two major limitations. First, only axial resolution can be improved. Second, mismatch between the spectral estimation model and spectrum data will cause inaccurate intensity reconstruction. For example, the estimated signal intensity fluctuated in the auto-regressive method. Spurious peaks appeared in the image when an inappropriate model order was chosen [21]. Similar problems were also observed with the maximum entropy model and the Prony model [20,22].

The deconvolution method can achieve both axial and lateral resolution improvement at a lower computational cost with the Wiener filtering method and the Lucy-Richardson (LR) deconvolution method being the most notable examples [23-25]. These methods acquired the point spread function (PSF) of OCT system to realize the resolution enhancement. Sufficient iterations of deconvolution operation can fully recover the high-frequency information of the image if no noise is present. However, images are often degraded by noise, which will cause the convergence to noise dominated solution after certain iterations, such as in LR deconvolution [26,27]. Therefore, it is usually recommended to stop reconstruction after a small number of iterations to avoid excessive image artifacts.

Recently, deep learning methods have shown strong capability in OCT image resolution improvement [28-30]. For example, conditional generative adversarial networks (cGANs) was explored to enhance the optical axial and lateral resolution of OCT images while preserving and improving the detail of speckle content [29]. High resolution OCT images with about 1 μm isotropic resolution were used as the ground truth paired with low-resolution image obtained by synthetically resolution degradation. However, the model performance was sensitive to the input image size. Deep learning-based digital refocusing was also reported to extend OCT depth-of-focus and improve the image lateral resolution [30]. Deep learning methods have provided many new ideas for OCT resolution improvement. However, the datasets with correct low-resolution and high-resolution mapping are critical for training neural network models. Meanwhile it generally requires a large data size to produce realistic super resolution OCT images.

Prior knowledge as important input has been used for various OCT image processing, demonstrating clear effectiveness [31-35]. Among them, the sparsity and continuity prior knowledge were widely used to preserve the images texture and reduce OCT speckle noise. For instance, sparsity-based and segmentation-based sparse reconstruction methods for high image digital resolution were proposed [32,33]. Shearlet-based total variation de-speckling framework method was proposed by introducing a de-speckling continuity prior into the standard total variation model to achieve simultaneous noise reduction and texture detail preservation [34].

In this work, inspired by the introduction of prior knowledge, we propose the sparse continuous deconvolution algorithm with both sparsity and continuity as the prior knowledge to constrain the iterative deconvolution process. Since the total information carried by the low-resolution OCT image is invariant, adding a prior knowledge may help reveal more details of the object. We transfer the LR deconvolution process to an optimization problem and introduce the prior knowledge for it. The sparsity is used to ensure the high frequency information of the images through the resolution preserving regularization term. The continuity based on the correlation of the grayscale values is used to mitigate excessive image sparsity through continuous regularization term. Bregman splitting technique is then adopted to solve the resulting optimization problem. Axial and lateral resolution simulation study under different noise level were performed. The proposed method showed clear effectiveness. Optimal parameter selection was also studied to provide a general guidance. Further evaluation study on phantoms and biological samples show improved image resolution. It achieved nearly twofold resolution improvement for phantom beads image that can be quantitatively evaluated.

## 2. Theory

For a given imaging system, image can be regarded as the convolution of the object with the system PSF. For simplicity, OCT image can be expressed as:

$$f = N\{A \otimes x\} \tag{1}$$

where $f$ is the OCT image system produces, $x$ is the object and $A$ is the PSF of OCT system, and $N\{\ \}$ denotes the noise of OCT system.

### 2.1 Lucy-Richardson deconvolution

Lucy-Richardson deconvolution treats the recovery problem of the underlying image blurred by a known PSF as a statistical estimation rather than a direct inverse solution [23,25]. The iterative deconvolution is based on the Bayes' theorem of conditional probabilities. Eq. (2) gives the basic algorithmic computation:

$$g^{n+1}(x,y) = g^n(x,y)\left[\left(\frac{f(x,y)}{h(x,y) \otimes g^n(x,y)}\right) \otimes h(-x,-y)\right]. \tag{2}$$

$g^n(x,y)$ is the estimate of the undistorted image in the $n^{th}$ iteration. The optimization process starts with $g^0(x,y) = f(x,y)$ and interatively modifies $g^n(x,y)$ based on the PSF of the imaging system $h(x,y)$ and the original image $f(x,y)$.

Despite the theoretical feasibility of mathematical resolution improvement, LR method was usually unstable due to the presence of noise [26,27]. De-convolving the object from noise corrupted images usually confers an ill-posed inverse problem. Since the ultimate goal is to decode the reconstruction signal as close to their ground-truth as possible, we build up a constraint model using the widely shared prior knowledge, the sparsity and continuity, for OCT images, which are common features for biological samples.

### 2.2. Sparsity and Continuity in OCT image

The concept of sparsity in image can be visualized in Fig. 1(a) where the central point is of clear boundary between adjacent neighborhood. Since more detailed structures corresponding to clearer structural boundaries can be visualized in higher resolution image, resolution improvement can result in sparsity increase in OCT image. We introduce the sparsity as the prior knowledge to antagonize OCT image resolution degradation. Similar to the strategy adopted by compressive sensing signal processing, $l_1$ norm instead of the $l_0$ norm is used to calculate the image sparsity to reduce the computation complexity [36-39].

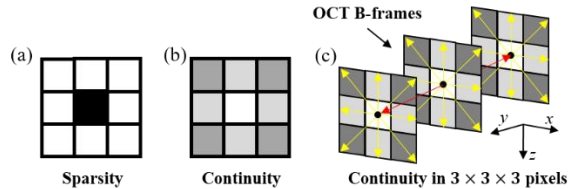

**Fig.1.** The concept of sparsity and continuity and OCT 3D continuity. (a) Specific examples for absolute sparsity and relative continuity (b). (c) Visualization of OCT 3D continuity in $3 \times 3 \times 3$ pixels and spatial coordinates in the world coordinate system corresponding to the OCT B-frames. yellow lines and red lines indicated the continuity of the center point.

However, if only the image sparsity is considered, some noise corrupted areas will produce unexpected over sharpening. The continuity prior is introduced to balance the sparsity. Biological samples are usually continuously changing, which corresponds to a certain correlation of image grayscale values within the 8-neighborhood region of each point in OCT images, as shown in Fig.1(b). Ensuring the correlation of OCT smoothed regions could avoid excessive sparsity caused over-sharpening. Thus, we introduced the continuity of a 3 ×3 region in OCT B-frame to mitigate this over sharpening. As for 3D OCT images, we also used the information between OCT adjacent frames to optimize the 3D spatial image continuity.

We used the continuity in $z$, $x$ axis (OCT B-mode image axis) and also $y$ axis (3D scanning direction) to suppress noise and subsequent reconstruction artifact. The 3D OCT continuity in $3 \times 3 \times 3$ pixels was presented in Fig.1(c). The yellow lines indicated the continuity correlation between the center point and its 8-neighboring points in different directions. And the red lines indicated the continuity between the central point and its adjacent frames.

In sparse reconstruction, image continuity is usually achieved by a constant first-order partial derivatives of the image. However, this traditional total variation penalty might over-sharpen the boundary between different regions, which would result in staircase-like and aberrant reconstructed images [40,41]. In our method, we used a second-order partial derivative penalty, denoted the continuous penalty, which focused on the piecewise-approximation of boundaries between regions of different intensities. It enabled global smooth transitions in the final reconstructed super resolution OCT images.

The continuity matrix regarding the structural continuity is defined as following:

$$R(f) = \left\| \begin{array}{ccc} f_{zz} & f_{zx} & \sqrt{\varepsilon_y} f_{zy} \\ f_{xz} & f_{xx} & \sqrt{\varepsilon_y} f_{xy} \\ \sqrt{\varepsilon_y} f_{yz} & \sqrt{\varepsilon_y} f_{yx} & \varepsilon_y f_{yy} \end{array} \right\|_1$$

$$= \|f_{zz}\|_1 + \|f_{xx}\|_1 + \varepsilon_y \|f_{yy}\|_1 + 2\|f_{zx}\|_1 + 2\sqrt{\varepsilon_y}\|f_{zy}\|_1 + 2\sqrt{\varepsilon_y}\|f_{xy}\|_1 \quad (3)$$

In Eq. (3), $f$ represents the non-optimized image. $\|*\|_1$ is the $l_1$ norm, $\varepsilon_y$ denote the regularization parameters that represents the continuity along the different $y$ axis. $\varepsilon_y$ is set to zero if the input data is only a 2D B-mode image. The subscript of $f$ in the matrix indicate the continuity along different axes.

In summary, we can see that it is essential to preserve both sparsity and continuity of the image. Here, we propose the sparse continuous deconvolution algorithm. We believe both continuity and sparsity prior knowledge are general features of the OCT images. These general features can suppress the image noise and facilitate high resolution information extraction collaboratively.

## 3. Methods

### 3.1 Derivation of algorithms

The loss function is shown in Eq. (4). The continuous matrix is used to reduce artifact and increase robustness at the price of reduced resolution. And the sparsity is used to balance the extraction of high-frequency information of resolution. which gives:

$$\arg\min_x \left\{ \frac{\lambda}{2} \|f - (A \otimes x)\|_2^2 + R(x) + \lambda_s \|x\|_1 \right\}. \quad (4)$$

The first term of the equation is the fidelity term, representing the distance between the recovered image $x$ and the original OCT image $f$. $A$ is the PSF of OCT system. The second and third terms are the continuity and sparsity priors. $\|*\|_1$ and $\|*\|_2$ are the $l_1$ and $l_2$ norms, respectively. $\lambda$ and $\lambda_s$ denote weight factors balancing the images' fidelity with sparsity. By adjusting the relationship between the fidelity term and the continuity term with $\lambda$, we can tune the use of image continuity. Equally, adjusting the relationship between the continuity term and the sparsity term with $\lambda_s$, image sparsity can be tuned [34].

To simplify the computation complexity, we introduce the intermediate variable $g$ for iterative calculation. This transforms Eq. (4) into a two-step optimization [42]:

$$g = \arg\min_{g} \left\{ \frac{\lambda}{2} \| f - g \|_2^2 + R(g) + \lambda_s \| g \|_1 \right\} \quad (5)$$

$$x = \arg\min_{x} \left\{ \| g - (A \otimes x) \|_2^2 \right\} \quad (6)$$

The solution of Eq. (5) can be translated into a convex optimization problem. To do that, we adopted the split Bregman algorithm, which is widely used in total variation problems due to its fast convergence speed and whose detailed information can be found in [42].

### 3.2 Pseudo-code Chart of Sparse Continuous Reconstruction

The implementation of sparse continuous reconstruction algorithm is summarized below.

**Algorithm 1.** Sparse Continuous Deconvolution
**Input:** Original OCT image $f$, PSF of OCT system $A$.
**Initialization:**
  1) Set an initial image $f = g^1$
  2) Set parameters: iteration numbers $N_k$, regularization parameter, etc.

**Iteration:**
  1) Solve the problem (5) via Split Bregman with $N_k$ as the iteration number to get g
  2) Solve minimization problem (6) via Lucy Richardson deconvolution to get final $x$

**Output:** High resolution image $x$

Where $N_k$ is the number of iterations, which we set as 50 in this work. And we used the method in [25] to calculate the PSF of our OCT system. The PSF was estimated by imaging phantom with small particles embedded in the agarose. We found the reliable spots in phantom images and fitted them to the 2D Gaussian function to extract the system PSF.

### 3.3 Experimental Setup

The measurements were performed with our homebuilt spectral-domain OCT system. The light source consisted of a super-luminescent diode with a central wavelength of 1310 nm with the FWHM of 60 nm. We set the image size to be 1024×500 (axial×lateral). The axial resolution was measured to be 12.6 μm in air, and lateral resolution was experimentally determined to be 16.7 μm near the focal plane. For better visual appearance, the images in the experiment were cropped. In the experiment with polystyrene beads samples, we selected an area of 200×400 size near the focal plane. And in the biological OCT samples experiments, we cropped the image size to 500×500.

Our process was performed on a personal computer with Intel Core i7-9700k CPU (3.6 GHz), Windows-10 64-bit operation system. The LR deconvolution method and our sparse continuous deconvolution method were performed in MATLAB (R2019b).

### 3.4 Quantitative evaluation

Four commonly used quantitative metrics are adopted to evaluate the performance of the proposed sparse continuous deconvolution. The signal-to-noise ratio (SNR), contrast-to-noise ratio (CNR), peak-signal-to-noise-ratio (PSNR) and structural similarity index measure (SSIM) are calculated to assess the efficiency of all compared methods [43,44]. Since SSIM requires the reference image, it is only used for the simulation study here. To perform quantitative evaluation of reference-free images, we used the sharpness metric function in [45] for quantitative analysis.

SNR is defined as the ratio of mean intensity in a foreground region containing structure (e.g., the red box region in Figure 3(d)) to the standard deviation of intensity in a background region (e.g., the yellow box region in Figure 3(d)):

$$SNR = 10\log_{10}\frac{\mu_f}{\sigma_b}, \tag{7}$$

where $\mu_f$ and $\sigma_b$ represent the mean intensity of a foreground region and standard deviation of intensity in a background region.

The contrast between the foreground regions and the background noise is measured using the CNR metric:

$$CNR = \frac{|\mu_f - \mu_b|}{\sqrt{0.5 \times (\sigma_f^2 + \sigma_b^2)}}, \tag{8}$$

where $\mu_f$ and $\mu_b$ are the mean intensity of the foreground and background, $\sigma_f$ and $\sigma_b$ are the standard deviation of the foreground and background, respectively.

Peak-signal-to-noise ratio (PSNR):

$$PSNR = 10\log_{10}(\frac{\max(I_f)^2}{\sigma_b^2}), \tag{9}$$

where $\max(I_f)$ denotes the maximal foreground intensity and $\sigma_b$ denotes the standard deviation of the background.

The sharpness metric function M is defined to be one divided by the total number of points in the axial scan intensity which are above a predetermined threshold [45]. The higher the image resolution, the larger the M value of the image. On the contrary, the lower the resolution, the smaller the M value.

## 4. Results and Discussion

### 4.1 Numerical Simulation Study

We first synthesized a ground-truth image containing eight parallel paired lines departed by different distances ranging from 5 μm to 40 μm. We set the intensities of lines to be 100% (first row), 75% (second row), and 50% (third row), relative to the maximal intensity of ground-truth, respectively. To simulate possible image degradation, the ground-truth was convoluted with a 20 μm FWHM PSF and degraded with OCT system noise plus 25% (first column), 50% (second column) and 75% (third column) Gaussian noise. The OCT system noise was obtained from air imaging. After that, we de-convolved these raw images (left, labeled as 'Raw') with LR deconvolution algorithm (middle, labeled as 'LR') and sparse continuous deconvolution (right, labeled as 'SC'), shown in Fig.2.

From Fig.2, we can see that the sparse continuous deconvolution has the ability of improving the axial resolution under different noise and intensity levels. And the conventional LR deconvolution covered greatly amplified artifact of the reconstructed images. According to Fig.2(a)-(c), LR deconvolution stopped resolving 20 μm lines at 25% Gaussian noise. While the sparse continuous image could separate 15 μm line pair. Compared to the sparse continuous image with 100% intensity, lower intensities images had lower contrast. And the parallel linear structure was disrupted by alternated bright and dim speckles. As for second column images, the LR images raised more artifact in case of increased noise shown in Fig.2(d). In particular, when the image intensity got lower, the artifact in LR images seriously affected the reconstructed linear structure shown in Fig.2(e)-(f). However, the sparse continuous images can still reach 15~20 μm resolution. And for third column images with 75% Gaussian noise, the sparse continuous images still extracted high resolution information. But we can no longer separated the line structure from LR images shown in Fig.2(g)-(i).

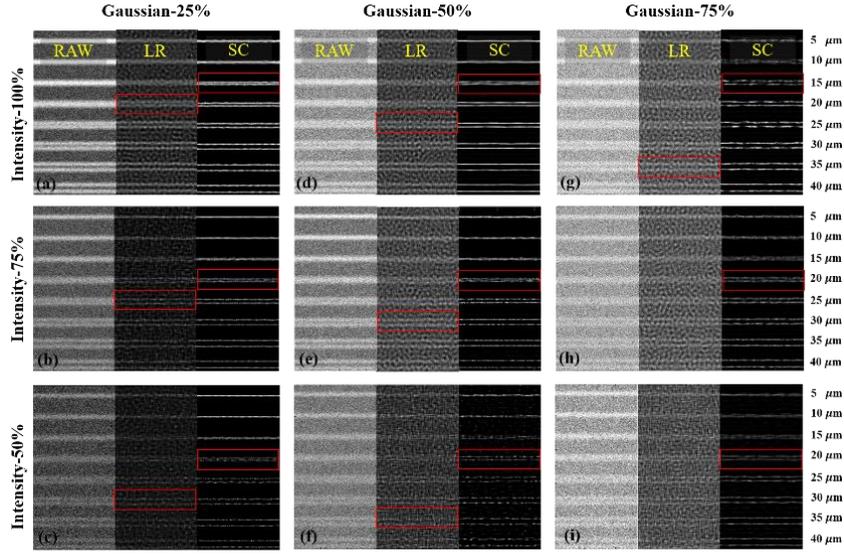

**Fig.2.** Effects of the combined noise on the paired lines resolved by the LR deconvolution and sparse continuous deconvolution. The red boxes mark the minimal resolution in different images.

To test the effect on irregular structures, four pairs of bisected-whole rings with 30 μm radius and 40/20/10/0 μm apart shown in Fig.3(a) were synthesized and convolved with a 20 μm PSF, and corrupted with OCT system noise plus 5% Gaussian noise shown in Fig.3(b). The conventional LR image shown in Fig. 3(c) shows excessive artifact. In contrast, although results of sparse deconvolution in Fig.3(d) did not faithfully generate identical images like ground-truth, it suppressed periodic artifact and resolved irregular degraded rings.

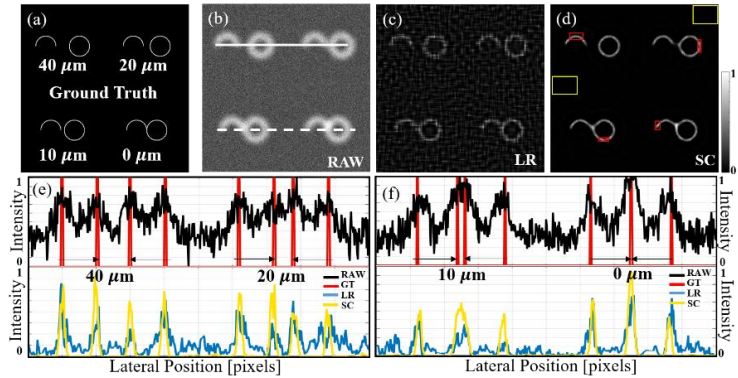

**Fig.3.** Partial ring simulation. (a) The ground truth. (b) The ground-truth convolved with PSF and corrupted noise. (c) LR reconstructed image. (d) Sparse continuous image. (e) Zoomed line images for (a) (b) (c) and (d) of solid line position in Fig.3(b). (f) Zoomed line images of dashed line position in Fig.3(b). The black, red, blue and yellow lines indicated the lines of the raw image, ground truth, LR image and the sparse continuous image, respectively.

From Fig.3(e), we can notice that the sparse continuous line (yellow line) has narrower FWHM than the LR line (blue line). Furthermore, the sparse continuous line almost removed the noise and restored the same background as the ground truth, while the LR line contains abundant noise.

Quantitative SNR, CNR, PSNR and SSIM comparison is shown in Table 1. We selected the red box area shown in the Fig.3(d) as the foreground region of the image, and the yellow box as the background region. The results show that the sparse continuous deconvolution performed better in each metrics. Nevertheless, the performance improvement comes at the cost of computation time. Thus, combining Table 1 with Fig.2 and Fig.3, we can see that sparse continuous algorithm reconstructing with the sparse and the continuity prior can reduce image noise, suppress artifacts and increase resolution.

Table 1. Comparison of the SNR, CNR, PSNR, SSIM results and running time for conventional Lucy-Richardson deconvolution and sparse continuous deconvolution of Fig.3.

|          | RAW          | LR           | SC           |
|----------|--------------|--------------|--------------|
| SNR (dB) | 8.241±1.62   | 7.034±1.81   | **14.562±2.05** |
| CNR      | 3.718±0.08   | 2.408±1.60   | **4.944±0.49**  |
| PSNR (dB)| 18.972±0.13  | 22.211±1.05  | **36.806±2.75** |
| SSIM     | 0.0007       | 0.0383       | **0.7264**      |
| Time (s) | NA           | **1.323**    | 3.397        |

*4.2 Optimal values of the sparsity and the fidelity*

As a computational super resolution method, sparse continuous deconvolution also faces caveats for optimal parameter tuning. Among content parameters, we usually set the iterative deconvolution number as a constant of 15 for the LR deconvolution method. Moreover, the correction of continuity and sparsity needed to be adjusted carefully. We discovered the key to the high-resolution images lay in the continuity and sparsity ratio through our tests.

To explore the effect of different ratios on reconstruction results, we synthesized ring-shaped structures corrupted with the noise of 5% and 25% amplitudes. We selected four different sets of $\lambda$ and $\lambda_s$ values to demonstrate the effect of ratios in Fig.4(a). We also systematically examined the reconstructed image SNR under different sets of $\lambda$ and $\lambda_s$ values with the SNR heat map shown in Fig.4(b).

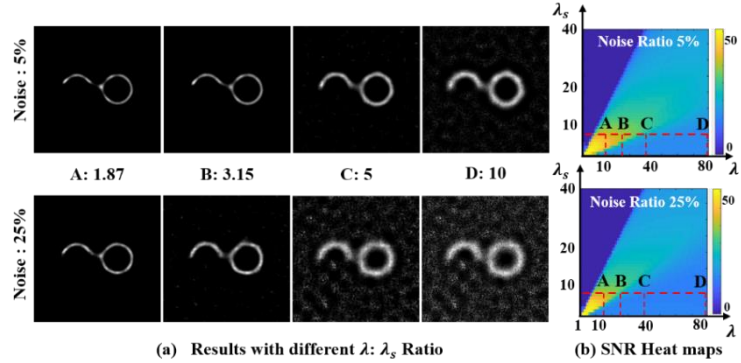

**Fig.4.** Exploring optimal $\lambda$ and $\lambda_s$ ratio choices for reconstructions of various noise images. (a) The reconstruction results of noise at 5% and 25% with four different ratios. The same column image represented the same ratio. Four different ratios were marked in (b).   (b) Reconstructed image SNR heat maps under different noise level.

Shown in Fig.4(a), with 5% noise, sparse continuous deconvolution almost resolved all the ring-shaped structures with high resolution. However, when the ratio was increased to 10, the image artifact appeared. With 35% noise, the ring became irregular when the ratio was greater than 5. Meanwhile, same ratio images under different noise levels reflected that the original image with high noise level required a ratio reduction to ensure the reconstructed image resolution.

From Fig.4, three key factors can be summarized. Frist, the $\lambda$ value must larger than $\lambda_s$ value. When the ratio was smaller than 1, the SNR of reconstructed images rapidly dropped to zero. Next, we found that the original images with high SNR afford larger ratio, while low SNR original images required small ratio. Finally, we suggest using small parameter values to ensure better image reconstruction SNR. It worth pointing out that SNR alone is not a good parameter to evaluate the image resolution improvement. Therefore, above factors can only serve as a general guidance instead of the standard.

*4.3 The polystyrene beads phantom*

To validate OCT image resolution improvement, we imaged a phantom of 10 μm polystyrene beads embedded in agarose gel. The results are shown in Fig. 5.

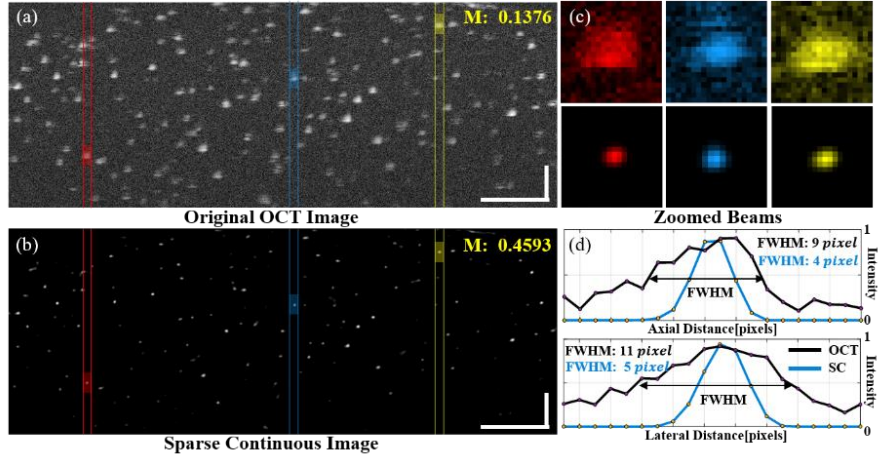

**Fig. 5.** Experimental validation of the resolution improvement of sparse continuous method on the polystyrene beads phantom image. (a) original image. (b) sparse continuous image. The sharpness metric function M value is marked at the top of images. (c) 20×20 pixels area for three reliable bead images. The red, blue and yellow boxes in the images includes the reliable spot for the transverse resolution assessment. (d) The average values of axial and lateral resolution line graph. Scale bar: 100 μm.

The image resolution is improved using the processed sparse continuous method in visual comparison. And the sharpness metric function value M of the images also shows the improvement of image resolution by sparse continuous method. Three representative beads region with 20×20 pixels are shown in Fig. 5(c). The corresponding axial and lateral profile are shown in Fig. 5(d). The sparse continuous method clearly improved both axial and lateral resolution, and it achieved a nearly two-fold axial resolution improvement and beyond two-fold lateral resolution improvement.

*4.4 The biological samples*

We further applied sparse continuous deconvolution algorithm to *ex-vivo* biological OCT samples. We first set the OCT 3D scan scanning range (along y axis) to 0 mm to obtain 20 identical B-mode images of orange sample. In this case we set the 3D parameter $\varepsilon_y$ to be 0, so that the algorithm is working in the 2D way. We then picked only one original OCT image as the input of algorithm to get single frame sparse continuous images.

Fig.6(a)-(c) presented a visual comparison of original single frame OCT image, single frame sparse continuous de-convoluted image and averaged original OCT image. Each of them was normalized. Zoomed region (red dashed box) in these images were marked and shown in Fig.6(d). Furthermore, the magnified line images corresponding to the vertical and horizontal position marked by the red dashed lines were plotted in Fig.6(e).

Visual comparison clearly showed the improved orange cell wall and cytoplasm contrast in single frame sparse continuous image compared with single frame OCT image shown in Fig.6(a)-(c). While the average method (Fig.6(c)) improved the SNR of single frame OCT image, it didn't improve the image resolution, which can be confirmed further from Fig. 6(e). Zoomed region in sparse continuous image was of higher resolution than the region in other images. As marked by the red arrows in Fig.6(d), the structure of the cell wall can be distinguished in sparse continuous area, while the OCT and averaged OCT areas cannot discriminate such detailed information.

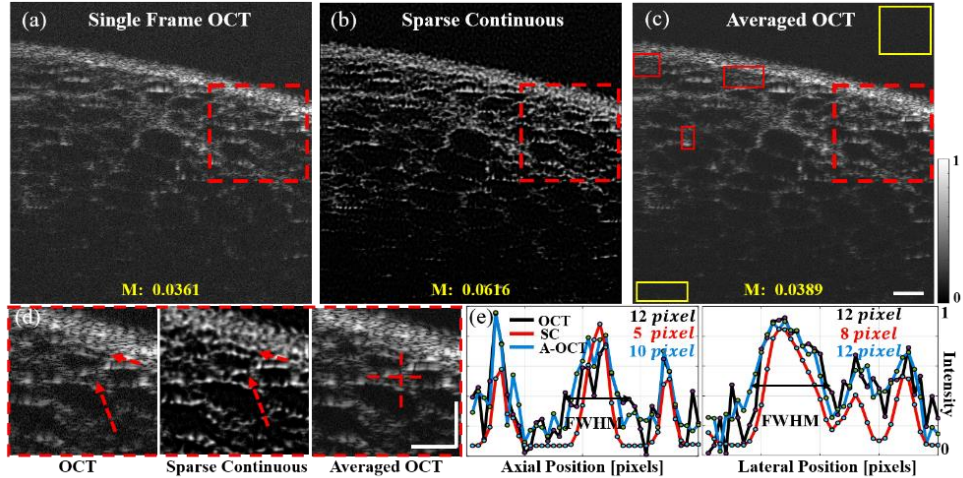

**Fig.6.** The compression of original OCT image, sparse continuous image and averaged OCT image on the orange sample. (a) single frame OCT image. (b) single frame sparse continuous image. (c) averaged OCT image. The sharpness metric function M value is marked at the top of images. (d) zoomed regions of red dashed boxes. (e) magnified line image marked in (d). The small red box area shown in the Fig.6(c) is the foreground region for Table 2, and the small yellow box is the background region. Scale bar: 200 μm.

Further, the magnified line image in Fig.6(e) exported the sparse continuous line (red line) had narrower FWHM with lower noise. The original OCT line (black line) contained high noise level and broader FWHM, which confirmed a lower resolution. We calculated and marked the FWHM of the signal peaks in Fig.6(e).

Compared with averaged OCT image, we found that the sparse continuous method didn't reconstruct the sample deep layers signal well. There might be two main reasons. First, the intensity of the signal degraded rapidly as the OCT imaging depth increase. In the deeper layers, signal was almost submerged in the background noise. Second, optimal parameter selection can lead to better image reconstruction. Fig.7 shows reconstruction images of the same orange sample with different $\lambda$ and $\lambda_s$ ratios. The artifact is produced with the ratio increased. Meanwhile, a too small ratio can cause the image detail information loss.

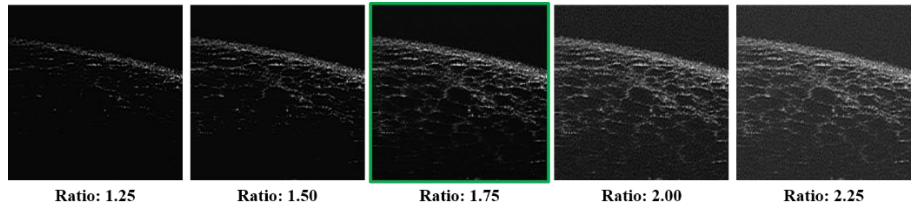

**Fig.7.** Different choices of $\lambda$ and $\lambda_s$ ratio for biological sample reconstructions. The image framed in green gives a good example of well-chosen ratio of Fig.6(a).

The quantitative SNR, CNR and PSNR comparison result is shown in Table 2. We selected the small red box area shown in the Fig.6(c) as the foreground region of the image, and the small yellow box as the background region.

As we expected, the single frame sparse continuous image showed higher evaluation metrics compared to single frame OCT images. Meanwhile, it also provided an improved SNR CNR and PSNR compared to averaged OCT image.

**Table 2. Comparison of the SNR, CNR and PSNR for original Single Frame OCT image, Single Frame Sparse Continuous image and averaged OCT image.**

|  | Single Frame SC | Single Frame OCT | Averaged OCT |
|---|---|---|---|
| **SNR (dB)** | **12.557±2.03** | 8.541±1.81 | 11.068±1.79 |
| **CNR** | **2.868±1.07** | 2.281±0.93 | 2.762±0.94 |
| **PSNR (dB)** | **32.301±1.84** | 23.424±1.41 | 29.231±1.54 |

We then set the OCT the 3D scan direction scanning range (along $y$ axis) to be 1 mm to obtain a volume OCT image of the orange sample with 256 frames. Meanwhile, we set the 3D parameter $\varepsilon_y$ to be 1. We then put all the OCT frames to sparse continuous deconvolution algorithm simultaneously. After that we can get the entire 3D volume result. This 3D method is different from the 2D method by using the continuity along the $y$-axis.

We show one selective B-mode image in 3D volume result and *enface* image at three different $y$ axis interval values. Among them, result of zero interval reconstructed 256 OCT B-mode images individually, which was the same as 2D reconstruction.

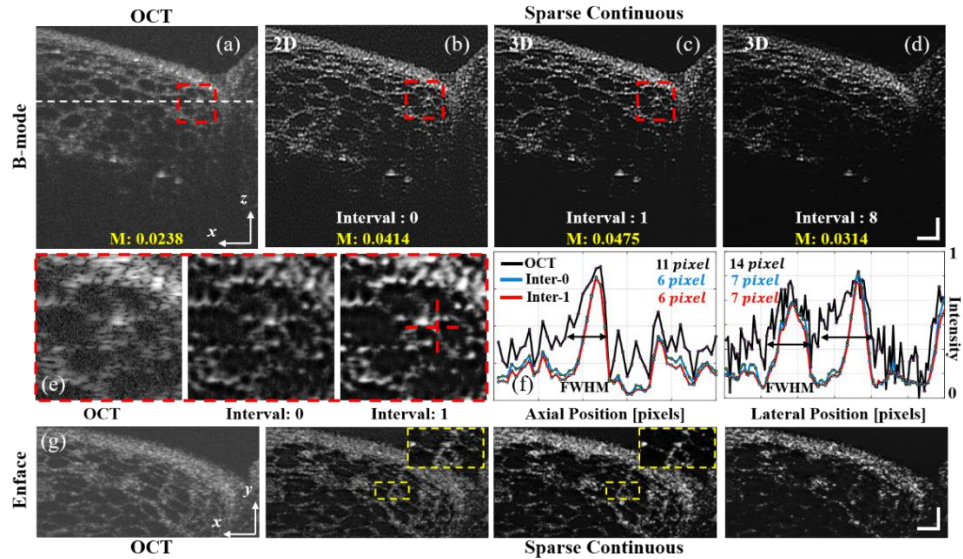

**Fig.8.** The orange OCT images comparison of original and sparse continuous reconstruction with different $y$ axis intervals. (a) original OCT B-mode image. (b) B-mode image of zero interval. (c) B-mode image for interval value of 1. (d) B-mode image for interval value of 8. The sharpness metric function M value is marked at the bottom of each image. (e) zoomed regions of red dashed boxes in (a)-(c). (f) magnified line image marked in (e). (g) The enface images of (a)-(d). Scale bar: 200 μm.

Both Fig.8 (b) with interval 0 and Fig.8 (c) with interval 1 achieved super resolution. Owing to the lack of adjacent frame continuity, some unexpected artifact was generated in Fig.8(b) with the same parameter set. Zoomed yellow regions in enface images also showed the artifact in Fig.8(b). Taking advantage of the object continuity between serial B-mode images, 3D sparse continuous reconstruction can provide better image quality shown in Fig.8(e). Quantitatively, the resolution of interval 0 (Fig.8b) and interval 1 (Fig.8c) achieved nearly twice improvement of original OCT image (Fig.8a), as shown in Fig.8(f). Meanwhile, the M value in different images also indicate that the resolution of sparse continuous reconstructed images is improved. However, when the interval value was set unreasonable large (interval 8), signal loss and image artifact will occur, as shown in Fig. 8(d).

Finally, we tested the 3D sparse continuous method on optical-cleared mouse spinal cord sample and *in vivo* human fingertip image volumes and the results are shown in Fig. 9 and Fig.10.

We can see consistent image quality and resolution improvement of the proposed method with interval 1. Quantitative comparison of M for original OCT image and sparse continuous with interval 0 and interval 1 also showed the resolution improvement.

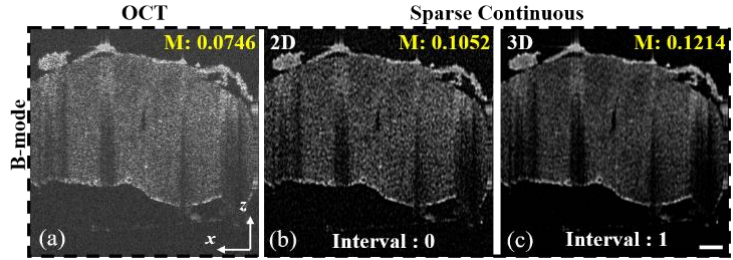

**Fig.9.** The original OCT B-mode image, 2D method reconstruction images and 3D method images with 1 interval value of optical-cleared mouse spinal cord sample. (a) Original OCT images. (b) 2D method reconstruction images with interval 0. (c) 3D method images with interval 1. The sharpness metric function M value is marked at the top of each image. Scale bar: 200 μm.

Comparing spinal cord and fingertip images with interval 1 and interval 0 in Fig.9 and Fig.10, we can find one interesting phenomenon, which is that there are speckle smooth effect of 3D sparse continuous reconstruction with interval 1. This indicates that continuity between adjacent frames is critical for 3D reconstruction as expected. We also find that the 3D fingertip sample has more speckle smooth effect than the 3D spinal cord sample. It shows a certain degree of image resolution reduction (M value drop) in 3D fingertip sample. The reason is that there is relative motion displacement between adjacent B-mode images since the fingertip was not still during the imaging process. Therefore, we registered the fingertip images to reduce motion effect of *in vivo* fingertip samples. We can see the M value improvement of registered-interval -1 over original B-mode image and 2D interval-0.

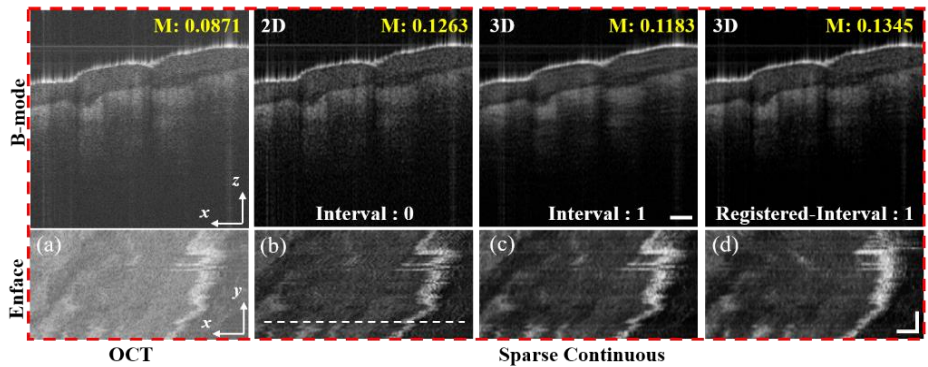

**Fig.10.** The comparison of *in vivo* fingertip sample images. (a) The original OCT images. (b) 2D method reconstruction images. (c) Unregistered 3D method images with 1 interval value. (d) Registered 3D method images. The corresponding position of each B-mode image is marked by white dashed line in enface image (b). The sharpness metric function M value is marked at the top of each image. Scale bar: 200 μm.

## 5. Conclusion and Future Work

In conclusion, we proposed the sparse continuous deconvolution method for OCT image resolution improvement by introducing the sparsity and continuity as the prior knowledge to constrain the iterative deconvolution. Both simulation study and experimental study on phantoms and biological samples have shown the advantages of our proposed method in terms of resolution, SNR, CNR and PSNR. Guidance over the parameter selection was provided. Both 2D and 3D OCT images can be processed with our method.

As a computational super resolution method, sparse continuous deconvolution also faces challenges associated with its forerunners. For example, in addition to resolution enhancement limited by the original image SNR, whether sparse deconvolution provided high-fidelity super resolution images also depended on the optimal selection of parameters. Experimenting with the sparse continuous parameter, we conclude that we need to adjust the fidelity and sparsity values carefully to achieve a well reconstructed OCT image.

Meanwhile, the algorithm improves image resolution by taking advantage of the sparsity and continuity of the samples. Erroneous reconstruction occurs when the continuity of the original samples is mismatched, which needs to be taken carefully in 3D reconstruction. A large gap between adjacent B-modes in 3D data will generate false continuity prior, which requires an increased sampling rate on *y* axis or data interpolation preprocessing. And *in vivo* samples also need to be pre-registered to reduce the motion effect of the sample.

Currently, the processing speed of the proposed algorithm is relatively slow for real time applications. However, super resolution imaging achieved by the deep learning algorithms can promise a significant speed-up to the entire processing. In the future, the combination of deep learning may help realizing the real-time processing. A good example can be found in [46]. Beyond that, designing automatic parameter tuning procedures may be another important work in the future. We expect our sparse continuous deconvolution method can be broadly tested to improve current OCT image resolution.

**Funding.** National Natural Science Foundation of China (62275023, 61505006); Beijing Institute of Technology (2018CX01018); Overseas Expertise Introduction Project for Discipline Innovation (B18005); CAST Innovation Foundation (2018QNRC001).

**Disclosures.** The authors declare that there are no conflicts of interest related to this article.

**Data availability.** Data underlying the results presented in this paper are not publicly available at this time but may be obtained from the corresponding author upon reasonable request.